\documentclass{ws-procs961x669}            

\usepackage[super,compress]{cite}
\usepackage{url}
\usepackage{siunitx}
\usepackage{hyperref}

\usepackage[utf8]{inputenc}	
\usepackage[T1]{fontenc} 	

\usepackage{aas_macros}

\begin{document}


\title{The Role of Self Interactions in the Cosmological Evolution of Warm Dark Matter
}

\author{R. Yunis,$^{1,2}$, C. R. Arg{\"u}elles,$^{1,3}$ D. L\'opez Nacir,$^{4,5}$ C. Sc\'occola$^{3,6}$, N. Mavromatos,$^{7,8}$ A. Krut$^{2}$}

\address{
$^{1}$International Center for Relativistic Astrophysics Network -- ICRANet,\\
Piazza della Repubblica 10, 65122 Pescara, Italy\\
$^{2}$ICRA, Dipartimento di Fisica, Sapienza Universit\`a di Roma, \\
Piazzale Aldo Moro 5, 00185 Rome, Italy\\
$^{3}$Facultad de Ciencias Astronómicas y Geof\'isicas, Universidad Nacional de La Plata,\\
Paseo del Bosque, B1900FWA La Plata, Argentina\\
$^{4}$Departamento de F\'isica Juan Jos\'e Giambiagi, FCEyN UBA\\
$^{5}$IFIBA CONICET-UBA, Facultad de Ciencias Exactas y Naturales, Ciudad Universitaria,
Pabell\'on I, 1428 Buenos Aires, Argentina\\
$^{6}$CONICET (Consejo Nacional de Investigaciones Cient\'ificas y T\'ecnicas), Argentina\\
$^{7}$Theoretical Particle Physics and Cosmology Group, Physics Department, King’s College London, Strand, London WC2R 2LS, UK\\
$^{8}$Physics Department, School of Applied Mathematical and Physical Sciences, National Technical University of Athens, 9, Heroon Polytechneiou Str., Zografou Campus, Athens 157 80, Greece.}

\begin{abstract}
In this work we present a summary of recent studies on the effects of elastic self interactions in the evolution of Warm Dark Matter models (WDM), focusing on structure formation and the evolution of cosmological perturbations. 
We pay special attention to a particular class of sterile neutrino WDM known as $\nu$MSM and provide examples for the case of vector field self interactions. 
We calculate the effects of assuming self interacting dark matter in X-Ray astrophysical observations, in the formation of fermionic DM halos in (quasi) equilibrium states and in the evolution of DM perturbations in the early universe, assuming particle masses between $\mathcal{O}(1-100)$ keV. In the latter topic, we perform simulations using a modification to the public Boltzmann solver CLASS and compare our results with observations. 
We find self interactions to be an interesting addition to WDM models, which can alleviate tensions both present in standard CDM cosmology and regarding WDM itself, as well as provide an interesting avenue for DM halo formation. 

\smallskip

Report Numbers: KCL-PH-TH/2021-83

\end{abstract}


\bodymatter


\section{Introduction}

%
%

Several observations at large and small scales, such as the distribution of large-scale structure, CMB anisotropies and the internal structure of DM halos has lead to a standard model of cosmology: $\Lambda$CDM\cite{Bahcall:1999xn,PLANCK2018}. 
DM in this model is assumed to be ``Cold'': a distribution of colissionless, non relativistic particles with negligible velocity dispersion, which forms structure in a ``bottom-up'' fashion. 
However, recent observations have challenged this paradigm \cite{Bullock2017}~; namely, the so called 
missing satellites problem\cite{Griffen2016,Janesh_2019,TheDESCollaboration2015a}~; 
the too big to fail problem\cite{Boylan-Kolchin2011a}~; 
and the core-cusp problem\cite{Oh2015,Navarro1997,Bullock2017}~, among others. 
While these by no means rule out the standard model, it has sparked interest in the community in other families of DM models. 

We consider particle DM candidates in one of these families of models, known as Warm Dark Matter (WDM)\cite{Bode2000,Lovell2012,Boyarsky2018}~,  where the DM particles are produced while relativistic, but become non relativistic before the energy budget of the universe is dominated by matter. 
These particles would have a significant amount of velocity dispersion today, which causes a significant reduction of small scale structure population and an alleviation to some of the tensions we mentioned above. 
While traditionally these models were also expected to aid the \textit{core-cusp} problem, it was shown\cite{Maccio2012} that the particle mass requirement for this to happen was inconsistent with existing bounds, in what was known as the ``catch-22'' problem in WDM.
However, in recent years it has been shown that WDM halos with a core-halo distribution can exist\cite{Arguelles2019} (and are indeed stable\cite{Arguelles2020a}) as end results of the process of violent relaxation, which fit observations from both Milky Way and dwarf galaxies. 
While these collisionless relaxation processes are thought to result in the formation and thermalization of these halos\cite{1967MNRAS.136..101L,Chavanis2019}~, it has been argued in\cite{Yunis2021} that the process of Self Interactions can contribute to this end.

A particularly interesting particle candidate that can belong in the family of WDM models can be found in $\nu$MSM (Neutrino Minimal Standard model)\cite{Adhikari2016,Boyarsky2018} which assumes a light sterile neutrino to be the sole DM component, produced out of equilibrium via neutrino oscillations in the plasma, and requiring only a minimal extension to the Standard Model Lagrangian.
Several observations heavily constrain the parameter space of this model, such as X-Ray constraints from astrophysical objects \cite{Perez2017,Cherry2017,Ng2019a}~, structure formation constraints \cite{Schneider2016,Enzi2020a}~, production bounds \cite{Venumadhav2015})~, among others.
In recent years, Lyman-$\alpha$ observations have been particularly important, almost entirely ruling out the non-thermal production of WDM sterile neutrinos \cite{Viel2013,Enzi2020a}. 
These tight constraints may be alleviated by performing simple extensions to this model: in particular, we focus here on the inclusion of DM Self Interactions, in what we name Self Interacting Warm Dark Matter (SIWDM)\cite{Yunis2020b}. 
As we will summarize in this work, these may not only contribute in the aforementioned challenges, but also allow us to potentially address tensions in $\Lambda$CDM previously unaccounted by WDM models alone. 

\section{Self Interactions and WDM}

The addition of DM Self Interactions is a well studied problem, with many realizations mostly based on CDM models\cite{2016MNRAS.460.1399V,Tulin2017}. 
Its main assumption is the existence of non gravitational interaction terms between DM particles through a Standard Model or a Dark mediator. 
Typically, in the realm of structure formation the interactions considered are elastic, and therefore do not change the number or identity of particles.
We focus here on interaction channels of these kinds in WDM and characterize their possible effects across cosmological history.

Self Interactions have been invoked in N-body simulations of structure formation, where it has been realized that the inclusion of these interactions can ``flatten'' the inner cores of dwarf galaxies (as indicated by observations) via a process of thermalization in the inner regions of these halos\cite{Almeida2021,Balberg2001}. 
Thus, a combined model of WDM with the addition of Self Interactions may overcome both structure formation (``missing satellites'' and ``too big to fail'') as well as inner halo dynamics (``core-cusp'') challenges at the same time, preserving the benefits of both WDM and Self Interacting models. 
While there can be many particle physics models that reflect these interactions, we focus here instead on maintaining a certain model independence: some examples of particular model realizations can be found in the literature\cite{Khlopov2013,Tulin2012,Aarssen2012,Buen_Abad_2015}. 

\begin{figure}
\centering
\includegraphics[width=1.0\hsize,clip]{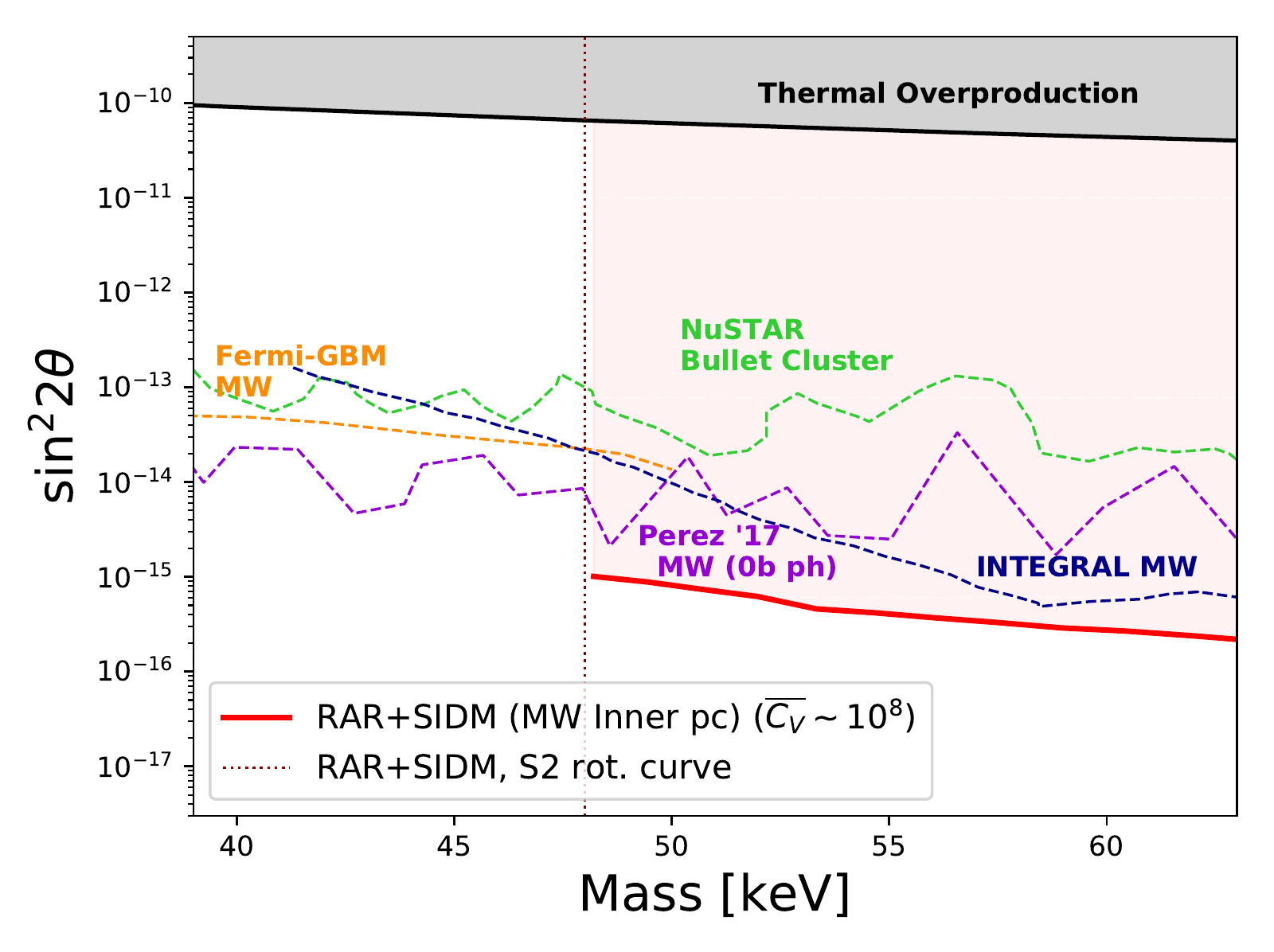}
\caption{Sterile neutrino parameter space limits obtained for MW GC observations using RAR$+$SIDM profiles (continuous red line), when assuming DM production due to self interactions and interaction strength according to Bullet Cluster constraints.
The light Red shaded region above the continuous red line corresponds to RAR$+$SIDM limits given by X-ray bounds (i.e. indirect detection analysis), while the vertical shaded region below 48 keV labels the smallest DM mass compatible with S-cluster stars' rotation curve data that can provide a BH alternative. 
The upper shaded region corresponds to production mechanism bounds, while other dotted lines refer to several X-ray bounds corresponding to different observations/DM profiles \cite{Perez2017}.
Reproduced from reference\cite{Yunis2020a}.}
\label{fig:RAR_vs_P2017}
\end{figure}

Since the first introduction of Self Interactions in WDM\cite{Hannestad00}~, it was realized that a combined model could provide interesting modifications in their dynamics. 
Indeed, Self Interactions do not only have an effect during structure formation, but also in perturbation evolution\cite{Huo2019,Egana-Ugrinovic2021} and DM production\cite{Boyarsky2018,deGouvea2019}. 
This last point was studied in the past in the context of $\nu$MSM and, in particular, as mentioned in reference\cite{Yunis2020a}, the phenomenon of mediator decay can significantly relax the parameter space of the model.
There it was considered the case of a WDM model with vector field interactions, previously suggested in the literature to provide interesting consequences for DM halos\cite{amrr}. 
These studies focus on WDM core-halo distributions\cite{Arguelles2019}~, known in the literature as RAR profiles, extended it via a Self Interacting model and calculated its potential effects on $\nu$MSM bounds coming from X-Ray observations. 
Interestingly, although bounds of the same order of magnitude as previous studies (that use N-body simulation-borne DM distributions) were obtained when considering observations of the MW galactic center, it was shown that these core-halo distributions are only compatible with $\nu$MSM bounds with the inclusion of additional production channels, such as Self Interactions. 
We present these results here in figure \ref{fig:RAR_vs_P2017}~, where we plot the parameter space for $\nu$MSM, together with the bounds obtained from X-Ray observations of the MW Galactic Center.

While it was shown in reference\cite{Yunis2020a} that the DM distribution of these halos is not affected by Self Interactions, it is an interesting open question if the formation of these systems may be affected by the inclusion of DM scattering. 
Indeed, these profiles are constructed assuming (General Relativistic) thermal equilibrium between DM particles. 
While coarse-grained equilibrium can be achieved by a collisionaless fluid, as was shown in previous works\cite{Arguelles2019,Chavanis2019,1998MNRAS.300..981C}~, Self Interactions can thermalize the inner regions of these systems\cite{Almeida2021} and, potentially, lead to the formation of a fermionic core. 

\section{SIWDM and Linear Cosmology}

Self Interactions have already been found to have significant effects in the evolution of linear cosmological perturbations\cite{Hannestad00,Huo2019,Egana-Ugrinovic2021}. 
In particular, the effects of these interactions can be realized in the power spectrum and, in the realm of WDM, it can lead to interesting scenarios such as non relativistic self decoupling (a.k.a. late kinetic decoupling). 
Particularly, the first challenge consists in implementing correctly a collision term in the first order Boltzmann equation for massive species\cite{Oldengott2014}. 
While many implementations involve performing fluid approximations for the DM component\cite{Heimersheim2020,Garny2018}~, it was realized that this approach is inaccurate in the case of light relics, such as SM interacting neutrinos\cite{Oldengott2014,Oldengott2017}. 
Based on this realization, some of us obtained a reduced, kernel-based form for the Boltzmann collision term\cite{Yunis2020b}~, based on an ansatz for the interaction amplitude that encompasses most tree-level massive mediator models in an exact way. 
We refer the reader to reference\cite{Yunis2020b} for a full expression of these terms. 

The system of equations derived in\cite{Yunis2020b} allows us to study the evolution of DM perturbations under the Relaxation Time Approximation (RTA)\cite{Krapivsky_2010}~, shown to be very precise for the case of massive mediator interactions\cite{Oldengott2017}~. 
We can express the Legendre expansion of the first order Boltzmann equation in the synchronous gauge as 

\begin{equation}
\mbox{\small $
\begin{split}
\dot{F_0} (k, E_q, \tau) \simeq& - \frac{q k}{E_q} F_1 (k, E_q, \tau ) + \frac{\dot{h}}{6} \frac{\partial f_0}{\partial \ln q}  \\
\dot{F_1} (k, E_q, \tau) \simeq& \frac{q k}{3 E_q} F_0 (k, E_q, \tau ) - \frac{2 q k}{3 E_q} F_2 (k, E_q, \tau ) \\
\dot{F_2} (k, E_q, \tau) \simeq& \frac{q k}{5 E_q} \Big[ 2 F_1 (k, E_q, \tau ) - 3 F_3 (k, E_q, \tau ) \Big] - \frac{\partial f_0}{\partial \ln q} \Bigg[ \frac{1}{15} \dot{h} + \frac{2}{5} \dot{\eta} \Bigg] - a \frac{F_2(k, E_q, \tau)}{\tau_{rel}} \\
\dot{F_l} (k, E_q, \tau) \simeq& \frac{q k}{(2l+1) E_q} \Big[ l F_{(l-1)} (k, E_q, \tau ) - (l+1) F_{(l+1)} (k, E_q, \tau ) \Big] - a \frac{F_l(k, E_q, \tau)}{\tau_{rel}} \\
& \hphantom{\frac{q k}{(2l+1) E_q} \Big[ l F_{(l-1)} (k, E_q, \tau ) - (l+1) F_{(l+1)} (k, E_q, \tau ) \Big] - a \frac{F_l(k, E_q, \tau)}{\tau_{rel}}} l \geq 3 ,
\end{split}
$}
\label{eq:RTA_Hierarchy_Motion}
\end{equation}

\noindent where $f(\vec{k}, \vec{q}, \tau) = F(\vec{k}, \vec{q}, \tau) + f_0(q,\tau)$ is the DM distribution, $F(\vec{k}, \vec{q}, \tau)$ is the first order perturbation and we have defined the Legendre moments and the gravitational potentials $h,\eta$ as in reference\cite{Ma1995}. 
The collision rate $\Gamma(\tau)$ is defined as 

\begin{equation}
\tau_{rel}^{-1} = \frac{g_i^3}{32 (2\pi)^3} \frac{\int dE_q \, dE_l \, ds \, f_{eq}(E_q, \tau)  \, f_{eq}(E_l, \tau) \, \chi(s)}{\int dE_q \, q \, E_q \, f_{eq}(E_q, \tau)} \ ,
\label{eq:RTA_TauDefinition_dE}
\end{equation}

\noindent where this expression is defined according to the notation in references\cite{Yunis2020b,Yunis2021}~. In the same references particular expressions for the collision kernel $\chi$ for various interaction models can be found. 

These equations involve the background DM distribution $f_0$. While in the standard approach, this function remains in a relativistic form reminiscent of WDM, a few circumstances can alter this assumption. 
In particular, if the Self Interactions maintain the DM fluid in kinetic equilibrium all the way until the fluid becomes non relativistic, the background at that moment will switch into a non relativistic form, constituting the scenario known as Non Relativistic Self Decoupling. 
The consequences of this scenario were explored by some of us in reference\cite{Yunis2021}. There, it was found that, if one imposes continuity of the limiting expressions for the energy density, the non relativistic distribution function becomes

\begin{equation}
f_0 (T<m) \sim 4.534 \mathcal{C}_{\rm NR} \exp \left[ - 1.075 q^2/ T_{\rm 0,R}^2 \right ] \ ,
\label{eq:NR_f0}
\end{equation}

\noindent where $\mathcal{C}_{\rm NR}, T_{\rm 0,R}$ are the normalization and temperature of the species in the relativistic limit. 
%

\section{SIWDM Cosmology: Simulations and Observation}

\begin{figure}[htb]
\centering
\includegraphics[width=1.0\hsize,clip]{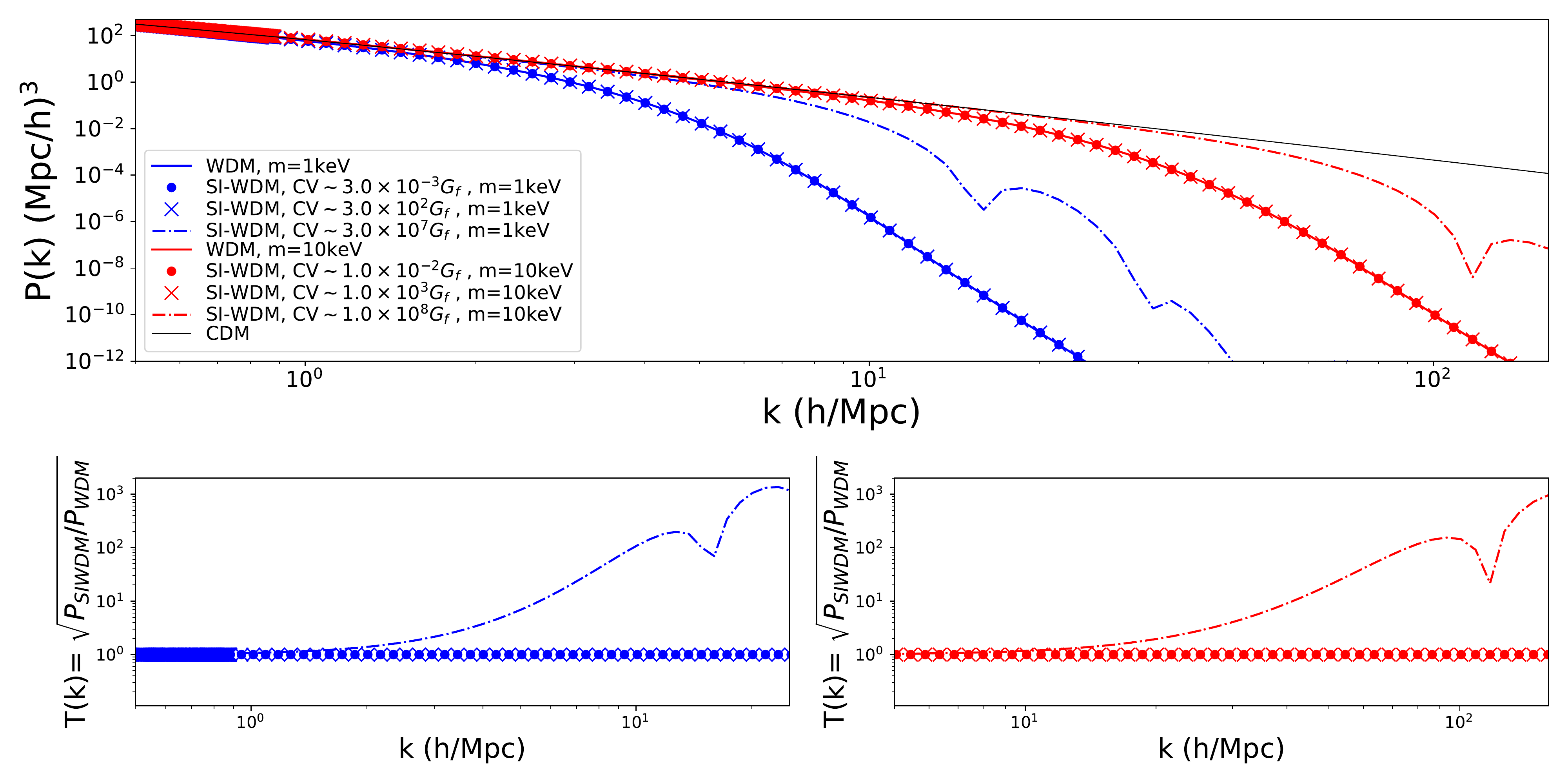}
\caption{
Power Spectrum (\textit{top panel}) and Transfer Functions with respect to standard WDM (\textit{bottom panels}) for a vector field SI-WDM model, simulated using a modification to CLASS.
We assume the relaxation time approximation (\ref{eq:RTA_Hierarchy_Motion}), and consider two values of the DM particle mass: $1$ and $10$ keV. Also plotted are the power spectra of a CDM model and of WDM models with DM mass of $1$ and $10\ {\rm keV}$. 
All WDM and SI-WDM models consider a nonresonant production scenario (Dodelson-Widrow mechanism, \cite{Dodelson1993a}) with $T \sim (4/11)^{1/3} T_\gamma$.
Reproduced from reference\cite{Yunis2021}
}
\label{fig:PowerSpectrum_NRSD}
\end{figure}

In reference\cite{Yunis2021} some of us developed a numerical implementation of these Self Interacting WDM models using a modified CLASS code, a publicly available Boltzmann solver \cite{CLASSIV}. 
This modification is available at the following link \href{https://github.com/yunis121/siwdm-class}{github.com/yunis121/siwdm-class}. 
We show there a few examples of the resulting power spectra for the Self Interacting models in figure \ref{fig:PowerSpectrum_NRSD}. 

\noindent There, we see some of the particular features of the models. We see that the inclusion of Self Interactions, while modifying slightly the high $k$ behavior of the power spectra, remain small elsewhere in the power spectrum for all models that do not undergo Non Relativistic Self Decoupling. 
However, for models with Non Relativistic Self Decoupling, the resulting power spectra may differ significantly from its relativistic counterpart. 
Indeed, we find that in this regime the models are ``colder'' (i. e. as if they correspond to a higher particle mass), and show even at smaller $k$ values a distinctive oscillatory pattern. 
This indeed has the effect of increasing the amount of small structure formed for these models. 

Most of the tensions inherent to $\nu$MSM WDM models come from structure formation, namely MW satellite counts and Lyman-$\alpha$ observations.
These are related to the fact that the preferred parameter ranges may underproduce small structure and almost rule out the available parameter space. 
So, the inclusion of Self Interactions can significantly relax the existing bounds on this family of models.

\begin{figure}[htbp]
\centering
\includegraphics[width=1.0\hsize,clip]{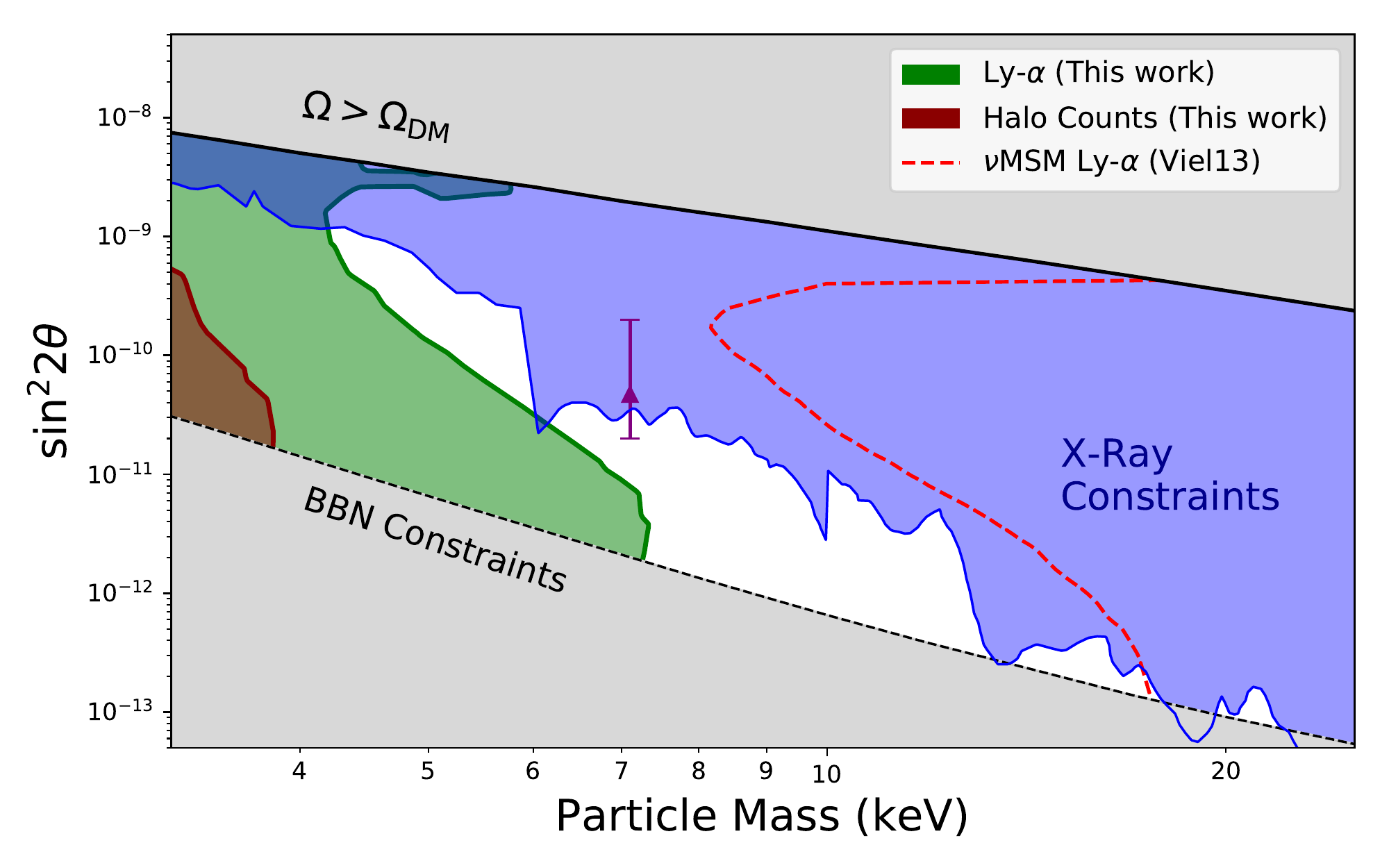}
\caption{\small 
Parameter space constraints for $\nu$MSM, where MW satellite halo counts and Lyman-$\alpha$ forest bounds are analyzed under a self interacting model as outlined above. 
We consider a self interacting model under a vector field mediator, with its interaction constant given by $\sigma / m  \sim 0.144 C_v^2 / m^3 = 0.1 \mathrm{ cm}^2/g$, the upper limit given by Bullet Cluster constraints \cite{amrr}.
For comparison, we plot the Lyman-alpha bounds for the non interacting case for a comparable analysis, according to the results in \cite{Viel2013}. 
We also plot other bounds to the $\nu$MSM parameter space for informative purposes \cite{Schneider2016,Cherry2017,Ng2019a,Boyarsky2018,Venumadhav2015}.
Also for informative purposes, we plot a model compatible with a tentative 3.5 keV DM signal as a purple triangle \cite{Bulbul2014,JeltemaProfumo,Cappelluti2018} for informative purposes.
Reproduced from reference\cite{Yunis2021}
}
\label{fig:Obs_Results_SInuMSM}
\end{figure}

Reference\cite{Yunis2021} provides an evaluation of the predictions of these models for the number of MW satellites\cite{Schneider2016} as well for the observations of the Lyman-$\alpha$ forest. 
We have found that a maximally interacting model (as allowed by the observations of the Bullet Cluster\cite{amrr}) can readmit a significant portion of the $\nu$MSM parameter space, and we illustrate this in figure \ref{fig:Obs_Results_SInuMSM}. 

\section{Conclusions}

We have studied in detail the evolution of Warm Dark Matter in the presence of Self Interactions.
With the aim of contributing to a detailed study of this subject from both a cosmological and astrophysical perspective we provide here an overview of the results obtained so far.

Reference\cite{Yunis2020a} presents and exploration of the consequences of considering WDM core halo RAR distributions\cite{Arguelles2019} in the parameter space of sterile neutrino WDM. 
There it is suggested that the inclusion of Self Interactions can at the same time relax the WDM sterile neutrino parameter space via additional production channels, readmit RAR core-halo distributions into these models and assist with relaxation and thermalization of these systems. 

Reference\cite{Yunis2020b} provides a theoretical framework for the treatment of cosmological perturbations in WDM that allows to quantify the effects of these extended models in the cosmological perturbations. 
There, we developed a kernel-based expression for the collision terms based on an ansatz for the interaction amplitude, as well as considered a few approximations to the resulting Boltzmann hierarchies such as the Relaxation Time Approximation. 
Reference\cite{Yunis2021} presents a numerical application to these models in CLASS and a comparison of the resulting spectra with MW subhalo and Lyman-$\alpha$ observables, both in the standard decoupling scenario as well as in the non relativistic case. 

We have reached conclusions that indicate the Self Interacting models may provide an interesting extension of the sterile neutrino WDM models. 
In the future, we hope to contribute further in the development of a more accurate treatment of these Self Interactions at a Cosmological level, as well as complementary models for early universe histories and, possibly, a deeper insight into the formation of these core-halo models. 
While further research is needed if this model is to be considered a viable alternative to standard DM approaches, we believe the work summarized here may be an important stepping stone in the study of these extensions. 

\section*{Acknowledgments}

CRA has been supported by CONICET, Secretary of Science and Technology of FCAG and UNLP (grants G140 and G175), National Agency for the Promotion of Science and Technology (ANPCyT) of Argentina (grant PICT-2018-03743) and ICRANet. 
CGS is supported by ANPCyT grant PICT-2016-0081; and grants G140, G157 and G175 from UNLP.
DLN has been supported by CONICET, ANPCyT and UBA.
The work of NEM is supported in part by the UK Science and Technology Facilities  research Council (STFC) under the research grant ST/T000759/1. 
NEM  also acknowledges participation in the COST Association Action CA18108 ``{\it Quantum Gravity Phenomenology in the Multimessenger Approach (QG-MM)}''.


\bibliographystyle{ws-procs961x669}
\bibliography{proceedings}

\end{document}